\documentclass[%
 reprint,
 amsmath,amssymb,
 aps,
]{revtex4-2}

\usepackage{graphicx}
\usepackage{dcolumn}
\usepackage{bm}

\usepackage{lineno}

\begin{document}

\preprint{APS/123-QED}

\title{Thermodynamic Origin of Degree-Day Scaling in Phase-Change Systems}

\author{Zhiang Xie}
 \email{xieza@sustech.edu.cn}
 \homepage{https://ymi33.github.io/en/}
\affiliation{%
 College of Ocean and Earth Sciences, Xiamen University 
}%
\affiliation{%
 Department of Earth and Space Science, Southrn University of Science and Technology 
}%

\date{\today}

\begin{abstract}
The temporal evolution of systems undergoing phase changes is constrained by thermodynamic thresholds, pinning state variables to fixed values. This threshold behavior redirects energy exchange into phase transitions, rendering the underlying energetics inaccessible from temperature alone and introducing strong nonlinearity. Here, we introduce the concept of latent temperature—a counterfactual thermal trajectory reconstructed by suppressing phase change, thereby recovering the underlying unconstrained evolution. We show that the latent trajectory is uniquely determined by a variational principle of minimum structural perturbation, minimizing nonlinear residuals relative to a linear relaxation process. The cumulative temperature elevation—mathematically equivalent to the one-dimensional Wasserstein distance between the latent and observed temperature distributions—is strictly proportional to the latent heat of phase change. The proportionality constant emerges naturally as the characteristic dissipation timescale $\tau$. This result provides a first-principles derivation of the classical Positive Degree-Day relationship and establishes phase-change thermodynamics as an optimal transport problem in non-equilibrium systems.
\end{abstract}

\maketitle

Thermodynamic state variables in driven dissipative systems are often subject to hard constraints imposed by phase transitions. Across diverse settings—from solid--liquid interfaces in materials science to the surface energy balance of the cryosphere—the observable temperature becomes pinned at the melting point during phase change, effectively collapsing the accessible thermal state space. While external forcing remains continuous, the temperature response is clipped, redirecting excess energy into latent heat. As a result, distinct energetic histories may project onto identical temperature trajectories, rendering temperature an incomplete descriptor of the system’s energetics. Although enthalpy-based formulations have been developed to ensure numerical closure of the total energy budget \cite{Idelsohn1994,Colera2025}, they do not resolve the inverse problem: inferring the underlying energetic evolution solely from a degenerate temperature record remains structurally ill-posed. 

A prominent yet heuristic circumvention of this degeneracy is the Positive Degree-Day (PDD) scaling \cite{Hock2005}, which remains a standard parameterization in Earth System Models for projecting cryospheric responses across diverse timescales \cite{Niu2019, Poppelmeier2023, Li2024}. To estimate melt rates, this empirical approach assumes that annual cumulative mass loss is proportional to the time-integrated daily air temperature above the freezing point \cite{Braithwaite1995}, effectively treating air temperature as a proxy for the energy input histories. While the PDD method exhibits remarkable predictive skill under relatively stable climatic regimes \cite{Zappa2003}, its lack of a first-principles derivation from the surface energy balance introduces structural uncertainties. This limitation becomes particularly acute under non-stationary climate conditions, where the historical coupling between air temperature and the surface energy budget may diverge, potentially rendering existing parameterizations unreliable for future projections \cite{Robinson2014}. A fundamental gap therefore persists between the rigorous physics of energy-balance formulations and the practical efficacy of temperature-index schemes.

In this Letter, we bridge this gap by introducing the \emph{latent temperature} $\theta(t)$—a counterfactual thermal trajectory describing the system’s evolution in an unconstrained state space where phase transitions are energetically suppressed. We show that energy conservation alone enforces an exact duality between the total latent heat absorbed during phase change and the cumulative exceedance of $\theta(t)$ above the melting threshold, mediated by a characteristic surface dissipation timescale. This relation is mathematically equivalent to the one-dimensional Wasserstein-1 distance \cite{figalli2021invitation} between the latent and observed temperature trajectories. Phase change thus emerges as an optimal transport process that projects continuous energetic variability onto a constrained thermodynamic boundary, providing a rigorous theoretical foundation for the PDD law.

To formalize this structure, we consider a generic system driven by a time-dependent energy flux. Once the melting threshold $\theta_f$ is reached, phase change imposes a nonlinear constraint on the thermal evolution: additional energy input no longer produces a temperature response but is sequestered as latent heat. This constraint acts as a rectifying operator, censoring thermal excursions above $\theta_f$ while preserving those below. The observed temperature trajectory $\theta_{\mathrm{obs}}(t)$ thus constitutes a projection of the system’s full energetic state onto a phase-limited thermodynamic manifold.

\begin{figure}
    \centering
    \includegraphics[width=\linewidth]{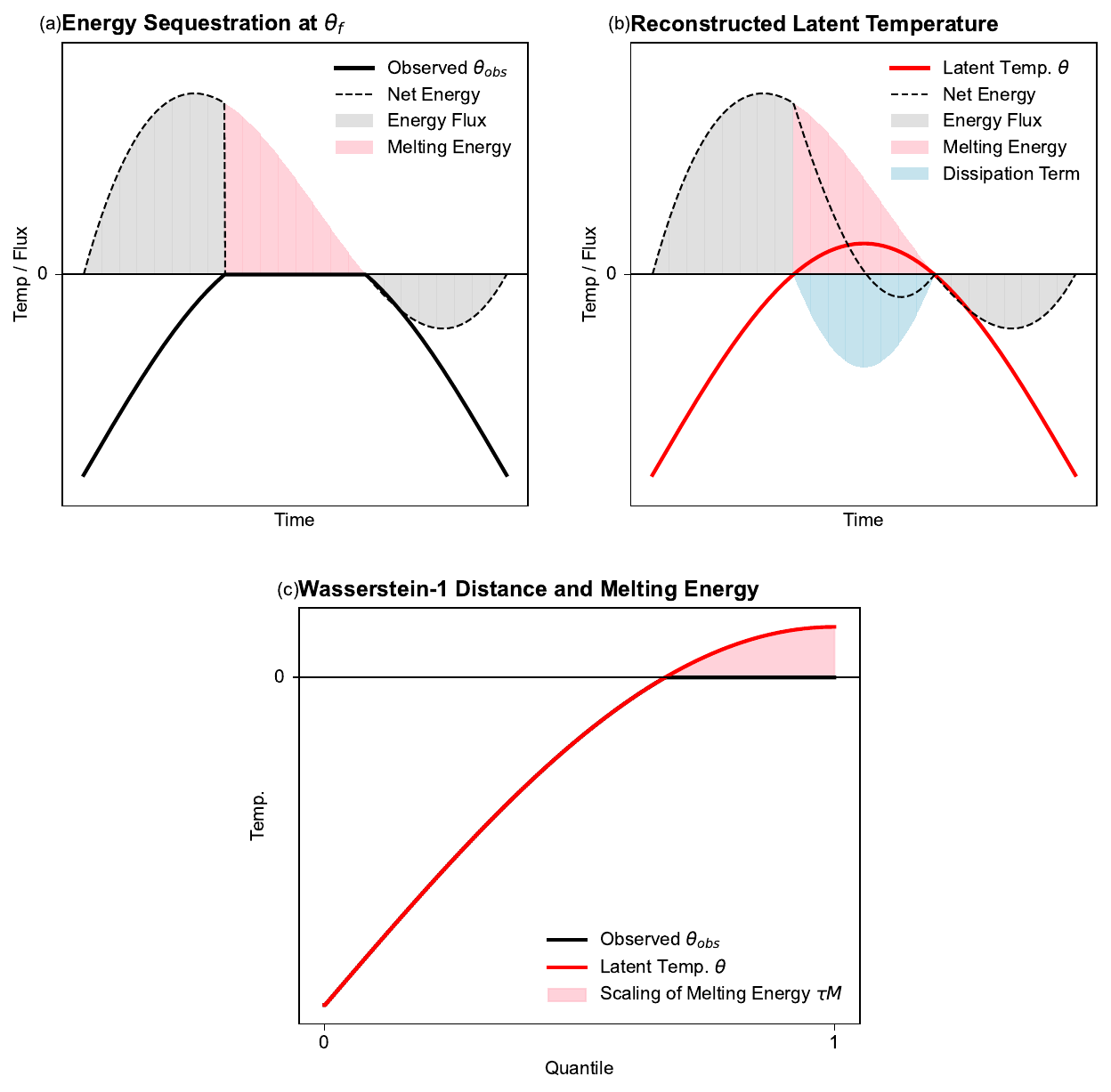}
    \caption{\textbf{Conceptual framework linking phase change and latent temperature.}
(a) In phase-change systems, temperature is constrained at the melting point $\theta_f$ (for convenience, $\theta_f=0$ in the diagram), rectifying continuous energy input into latent heat while suppressing observable thermal fluctuations.
(b) The latent temperature $\theta(t)$ denotes the counterfactual thermal trajectory the system would follow in the absence of phase change, encoding the energetics sequestered during melting.
(c) The cumulative exceedance of the latent temperature above $\theta_f$ is geometrically equivalent to the Wasserstein-1 distance between the latent and observed trajectories and scales linearly with the total melt energy $M$ via a characteristic dissipation timescale $\tau$.}
    \label{fig:diagram}
\end{figure}

During the melt season $t \in [t_s, t_e]$, the observed surface energy balance is governed by
\begin{equation}
\rho C_s \dot{\theta}_{\mathrm{obs}} = Q_{\mathrm{net}}(t) - Q_m(t),
\label{eq:obs_evo}
\end{equation}
where $Q_m$ denotes the latent heat flux consumed by melting (Figure \ref{fig:diagram}a). Since $\theta_{\mathrm{obs}}$ is pinned at the melting point ($\dot{\theta}_{\mathrm{obs}} = 0$), $Q_m$ effectively absorbs the net energy input. To reconstruct the unconstrained thermal state, we postulate that the latent heat flux can be decomposed into virtual storage, dissipation, and residual components:
\begin{equation}
    Q_m(t) = \rho C_s \dot{\theta} + \frac{1}{\tau}(\theta-\theta_f) + \mathcal{R}(t),
    \label{eq:lat_heat_decomp}
\end{equation}
where $\tau$ is a characteristic dissipation timescale and $\mathcal{R}(t)$ accounts for redistribution processes unresolved by the linear relaxation approximation. This decomposition does not introduce new physics but represents an exact bookkeeping of the latent heat flux under the phase constraint.

Substituting this decomposition into Eq.~(\ref{eq:obs_evo}) and enforcing the melt constraint yields the effective relaxation dynamics for the latent temperature:
\begin{equation}
\rho C_s \dot{\theta} + \frac{1}{\tau}(\theta-\theta_f) = Q_{\mathrm{net}}(t) - \mathcal{R}(t).
\label{eq:latent_dynamics}
\end{equation}
This formulation establishes a direct correspondence between the instantaneous surface energy balance and the latent state evolution. The latent temperature is thus physically identified as the dynamical variable whose relaxation compensates for the latent heat flux (Figure \ref{fig:diagram}b).

Thermodynamic consistency requires closure at the onset and termination of melting,
\begin{equation}
\theta(t_s)=\theta(t_e)=\theta_f ,
\label{eq:boundary}
\end{equation}
together with conservation of seasonal energy,
\begin{equation}
\int_{t_s}^{t_e} \mathcal{R}(t)\,dt = 0 .
\label{eq:energy}
\end{equation}
Equations~(\ref{eq:latent_dynamics})--(\ref{eq:energy}) define a constrained class of admissible latent trajectories, reflecting the intrinsic underdetermination introduced by phase change.

Integrating Eq.~(\ref{eq:latent_dynamics}) over the melt season and invoking the constraints (\ref{eq:boundary})--(\ref{eq:energy}) yields the identity
\begin{equation}
\int_{t_s}^{t_e} (\theta-\theta_f)\,dt = \tau M ,
\label{eq:pdd}
\end{equation}
where $M=\int_{t_s}^{t_e} Q_m\,dt$ is the total melt energy. Equation~(\ref{eq:pdd}) provides a rigorous thermodynamic foundation for the PDD relation: melt is strictly proportional to the time-integrated exceedance of the latent temperature, with $\tau$ acting as the proportionality coefficient (Figure \ref{fig:diagram}c). By combining the instantaneous decomposition [Eq.~(\ref{eq:lat_heat_decomp})] with the integral constraint [Eq.~(\ref{eq:pdd})], this framework establishes a dynamical bridge between surface energy balance formulations and degree-day scaling.

Geometrically, since $\theta_{\mathrm{obs}}(t)=\theta_f$ throughout the melt season, the left-hand side of Eq.~(\ref{eq:pdd}) corresponds to the $L^1$ distance between the latent and observed temperature trajectories. In one dimension, this distance is equivalent to the Wasserstein-1 metric $W_1(\theta,\theta_{obs})$ (see Appendix Section \ref{sec:w1_proof}). So 
\begin{equation}
W_1(\theta,\theta_{obs})  = \tau M .
\label{eq:wdis_pdd}
\end{equation}
Phase change may therefore be interpreted as an optimal transport process that compresses latent thermal fluctuations onto the melting boundary, with the melt energy determining the associated transport cost.

The identity (\ref{eq:pdd}) establishes $\tau$ as the proportionality factor linking latent thermal excursions to melt energy. To assess whether this timescale admits a concrete physical interpretation—and whether it aligns with established glaciological practice—we now examine its consistency with linearized surface energy balance processes near the melting point. This step is essential for connecting the abstract relaxation framework to the empirically calibrated PDD method widely used in ice-sheet modeling.

The relaxation timescale $\tau$ introduced in Eq.~(\ref{eq:latent_dynamics}) implies a specific physical mechanism for energy dissipation. Linearizing radiative and turbulent fluxes around the melting point $\theta_f$ gives
\begin{equation}
\rho_i C_s \frac{d\theta}{dt} \approx Q_i(t) - \left(C_{\rm sen} + 4\sigma \theta_f^3\right) \left[\theta(t)-\theta_f\right],
\end{equation}
where $C_{\rm sen}$ is the bulk aerodynamic coefficient for sensible heat flux, $\sigma$ is the Stefan--Boltzmann constant, and $Q_i$ is the incident net energy. Comparing this with Eq.~(\ref{eq:latent_dynamics}) reveals that $\tau$ is the inverse efficiency of surface heat loss:
\begin{equation}
\tau^{-1} \sim \frac{C_{\rm sen} + 4\sigma \theta_f^3}{\rho_i C_s}.
\end{equation}
Substituting this into the PDD relation (Eq.~\ref{eq:pdd}) and expressing melt energy as $M = \rho_i L_f a$ (where $a$ is ablation depth and $L_f$ is latent heat of fusion) yields the theoretical degree-day coefficient:
\begin{equation}
C_{\rm PDD} = \frac{1}{\tau \rho_i L_f} \approx \frac{C_{\rm sen} + 4\sigma \theta_f^3}{\rho_i L_f}.
\end{equation}

Using representative values for ice-sheet surfaces ($C_{\rm sen} \approx 22.5\,\mathrm{W\,m^{-2}\,K^{-1}}$, $\theta_f = 273.15\,\mathrm{K}$, $\sigma = 5.67\times10^{-8}\,\mathrm{W\,m^{-2}\,K^{-4}}$, $\rho_i = 917\,\mathrm{kg\,m^{-3}}$, and $L_f = 3.34\times10^5\,\mathrm{J\,kg^{-1}}$) \cite{Dommenget2011,Xie2022}, we obtain $C_{\rm PDD} \approx 7.6\,\mathrm{mm\,d^{-1}\,^\circ C^{-1}}$, well within the empirically inferred range ($1.5 - 8.5 \mathrm{mm\,d^{-1}\,^\circ C^{-1}}$ for Greenland)\cite{Wake2015}. This agreement indicates that the relaxation timescale $\tau$ emerging from the latent framework is quantitatively consistent with known radiative and turbulent dissipation processes.

\begin{figure}
    \centering
    \includegraphics[width=\linewidth]{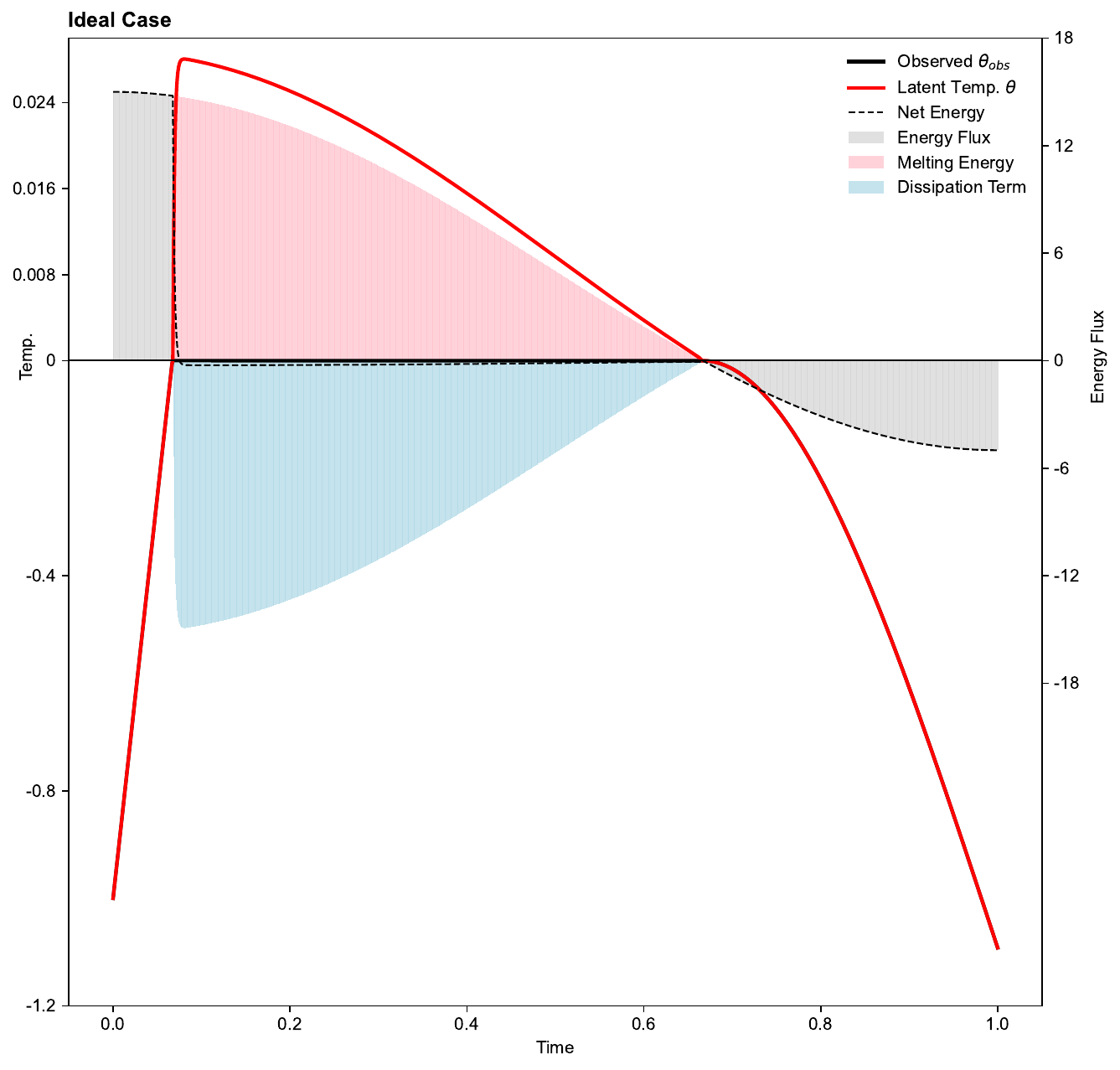}
    \caption{\textbf{Variational reconstruction of latent temperature under periodic forcing.} 
The system is driven by a biased sinusoidal energy flux $Q_i$ (gray and pink shading). 
The observed temperature $\theta_{\mathrm{obs}}$ (solid black) is strictly limited by the melting threshold $\theta_f=0$. 
To resolve the subtle energetic fluctuations hidden by this clipping, the vertical axis is split: 
the positive range ($[0, 0.024]$) is magnified relative to the negative range ($[-1.2, 0]$). 
The reconstructed latent temperature $\theta$ (red) recovers the unconstrained thermal state. 
The area representing sequestered melting energy (pink) is recovered through the integrated dissipation term (light blue), 
where the balance between these components defines the optimal relaxation timescale $\tau$. Black dash lines marks the net energy flux in reconstructed system. }
    \label{fig:ideal_case}
\end{figure}

To demonstrate the computability and physical meaning of the latent temperature framework, we consider an idealized periodic forcing experiment. The system is driven by a biased sinusoidal energy flux,
\begin{equation}
Q_i(t) = A \left[0.5 + \cos(\pi t)\right],
\end{equation}
with amplitude $A=10$. The function is chosen such that the positive forcing phase spans approximately two-thirds of the cycle to better display. For clarity, we adopt a normalized thermal inertia $\rho C_s = 1$ and reconstruct the latent temperature $\theta(t)$ subject to the energetic constraints described above and minimum $\mathcal{R}$ variation (see Appendix Section \ref{sec:supplementary_solution}).

Figure~\ref{fig:ideal_case} illustrates the reconstructed thermal and energetic evolution. During the initial cold phase ($t \lesssim 0.1$), the observed temperature $\theta_{\rm obs}$ (black) and the latent temperature $\theta$ (red) evolve identically. Upon reaching the melting threshold $\theta_f = 0$, phase change imposes a topological constraint that pins $\theta_{\rm obs}$ at the melting point, while the latent trajectory continues to evolve, encoding the energy sequestered during melting.

A salient feature of the reconstruction is the pronounced scale separation between the full-period temperature variability and the weak but energetically decisive excursions of $\theta$ above $\theta_f$. To reveal this hidden structure, the vertical axis employs a piecewise linear scaling that magnifies the melt-season interval. Within this window, the latent temperature exhibits a characteristic sequestration--relaxation profile, marked by a rapid rise at melt onset followed by a gradual decay as excess energy is dissipated.

Crucially, the reconstructed dissipation flux $\tau^{-1}(\theta - \theta_f)$ closely tracks the imposed net energy input $Q_i$ during melting, indicating that the latent temperature is not an arbitrary auxiliary variable but is dynamically conjugate to the latent heat flux. In this sense, $\theta$ acts as a fast-response variable that equilibrates external forcing against the phase-change constraint.

Quantitative analysis confirms exact seasonal energy closure. Energy accumulated during the melt window (gray bars) is precisely balanced by the reconstructed dissipation term (blue bars). The close correspondence in both magnitude and temporal structure between the melt flux $Q_m$ and the dissipation term implies instantaneous surface energy balance: 
\begin{equation}
    \tau^{-1}(\theta - \theta_f) \approx Q_m, \label{eq:inst_balance}
\end{equation}
consistent with the linear proportionality underlying the original PDD formulation \cite{Braithwaite1995}. Numerical integration further verifies the duality identity [Eq.~(\ref{eq:pdd})], with the inferred timescale $\tau$ agreeing with the analytical estimate from PDD relation to within $\sim 2\%$, attributable to discretization error. This idealized experiment establishes phase change as a rectifying operator that collapses continuous energetic variability onto the melting boundary, while the latent temperature provides a computable and thermodynamically complete representation of the hidden state space.

The framework developed here recasts phase-change thermodynamics as a constrained transport problem between latent and observable state spaces. By introducing the latent temperature as a counterfactual yet energetically faithful trajectory, we show that the PDD relation admits a first-principles interpretation rather than being purely empirical. Crucially, the proportionality between melt energy and integrated temperature exceedance emerges independently of the temporal structure of the forcing or the details of sub-seasonal energy redistribution. While the variational construction used in our idealized example illustrates the computability of the framework through minimal structural perturbation, the residual term $\mathcal{R}(t)$ admits a broader physical interpretation. In practice, $\mathcal{R}(t)$ represents the effective redistribution of melting energy and may be adapted to the dominant forcing mechanisms of the system. Importantly, the integral constraint on $\mathcal{R}(t)$ implies that its detailed temporal structure does not affect the validity of the PDD relation, but only modulates the instantaneous partitioning of energy.

This perspective provides a rigorous physical interpretation of classical semi-empirical glaciological models. Standard PDD schemes estimate melt from air-temperature exceedance, implicitly treating air temperature as a proxy for the latent thermal state $\theta$. Prior to melting, turbulent coupling maintains a close correspondence between surface and air temperatures; once phase change begins, this coupling is broken as the surface becomes pinned at the melting point while the air temperature continues to evolve. Our framework reveals that the PDD relation remains valid because the integral constraint on $\theta$ allows for significant freedom in the temporal distribution of the proxy variable (air temperature), provided the seasonal energy budget is closed. Consequently, the wide range of empirically inferred degree-day factors arises from the imperfect correspondence between the air temperature proxy and the true latent state $\theta$, modulated by the effective heat transfer efficiency encapsulated in $\tau$. The degree-day factor thus emerges as a physical transport coefficient rather than an arbitrary adjustable parameter. This interpretation predicts that degree-day factors should systematically covary with boundary-layer coupling efficiency and radiative damping, rather than being climate-invariant constants. 

Furthermore, by explicitly linking the PDD-type relaxation dynamics [Eq.~(\ref{eq:latent_dynamics})] to the instantaneous surface energy balance [Eq.~(\ref{eq:obs_evo})] via the decomposition of latent heat flux [Eq.~(\ref{eq:lat_heat_decomp})], we establish a unified framework that bridges the macroscopic scaling law with microscopic energetics. Introducing $\theta(t)$ effectively transforms the ill-posed inverse problem—inferring energetics from a degenerate clipped temperature record—into a structurally well-posed retrieval problem at the level of seasonal energetics. Since the relationship between the latent temperature rate $\dot{\theta}$ and the energy flux is bijective within this framework, $\theta$ serves as an energy-induced state variable that recovers the information lost during phase transition. This mapping resolves the apparent inconsistency between the time-integrated success of PDD schemes and the instantaneous complexity of the surface energy balance.

Beyond glaciology, these results point to a general principle for driven dissipative systems subject to threshold constraints. Whenever a state variable is pinned to a critical value while excess energy is diverted into a latent reservoir, the observable dynamics correspond to an optimal transport of suppressed fluctuations. Phase change therefore acts as a rectifying operator that projects unconstrained energetic variability onto a lower-dimensional manifold, with the associated transport cost set by an effective dissipation timescale. This viewpoint suggests natural extensions of degree-day concepts to other phase-limited systems, including permafrost thaw, sea-ice thermodynamics, and solid–liquid transitions in materials science.

\begin{acknowledgments}
The author would like to thank Dr. Dongwei Chen (Colorado State University) for inspiring discussions. This work was supported by the National Key Research and Development Program of China (No. 2024YFF0809001).
\end{acknowledgments}

\bibliography{apssamp}

\clearpage


\appendix



\section{Latent temperature and the 1-Wasserstein distance}
\label{sec:w1_proof}

Let $\theta(t)$ be a continuous temperature time series on $t\in[0,1]$, and let $\theta_{\mathrm{obs}}(t) = \min\{\theta(t),\theta_f\}$ denote the temperature clipped at the freezing point.

Because the integral of a function is invariant under reordering, we may rewrite:
\begin{equation}
    \int_0^1 \theta(t)\,dt
    = \int_0^1 F_\theta^{-1}(p)\,dp,
\end{equation}
where $F_\theta^{-1}$ is the quantile function of $\theta$. The same holds for $\theta_{\mathrm{obs}}$.

Since clipping only modifies the upper tail of the distribution, we have $F_\theta^{-1}(p) \ge F_{\theta_{\mathrm{obs}}}^{-1}(p)$ for all $p\in[0,1]$. Therefore:
\begin{equation}
    \int_0^1 \bigl[\theta(t)-\theta_{\mathrm{obs}}(t)\bigr]\,dt
    = \int_0^1 \left|F_\theta^{-1}(p)-F_{\theta_{\mathrm{obs}}}^{-1}(p)\right|\,dp.
\end{equation}
By definition, the right-hand side is the Wasserstein-1 distance between the two temperature distributions \cite{Santambrogio2015}. Hence, $W_l = W_1(\theta,\theta_{\mathrm{obs}})$.

In one dimension, the optimal Monge--Kantorovich transport map is given by this monotone rearrangement, independently of the temporal ordering of $\theta(t)$. This establishes that the latent temperature corresponds to the minimal redistribution work required to transform the unclipped temperature into its melt-limited counterpart, thereby identifying phase change as an optimal transport process in thermodynamic state space.

\section{Thermodynamic Origin of Degree-Day Scaling in Phase-Change Systems}
\label{sec:supplementary_solution}

\subsection{Governing Equation and General Solution}

We consider the latent temperature $\theta(t)$ governed during the melt interval
$t \in [t_s,t_e]$ by the linear relaxation model
\begin{equation}
\rho C_s \frac{d\theta}{dt}
+ \frac{1}{\tau}(\theta-\theta_f)
= Q(t) - \mathcal{R}(t),
\label{eq:s_ode}
\end{equation}
where $\tau$ is a characteristic dissipation timescale and $\mathcal{R}(t)$
represents a redistribution of energy in time accounting for unresolved processes.

Equation~(\ref{eq:s_ode}) admits the formal solution
\begin{equation}
\theta(t)
= \theta_f
+ e^{-\lambda (t-t_s)}
\int_{t_s}^{t}
\frac{Q(s)-\mathcal{R}(s)}{\rho C_s}
e^{\lambda (s-t_s)}\,ds ,
\label{eq:s_solution}
\end{equation}
where $\lambda = (\rho C_s \tau)^{-1}$.
The integration constant is fixed by the boundary condition
$\theta(t_s)=\theta_f$.

\subsection{Admissibility Constraints}

Thermodynamic consistency imposes two global constraints on $\mathcal{R}(t)$.
First, conservation of seasonal energy requires
\begin{equation}
\int_{t_s}^{t_e} \mathcal{R}(t)\,dt = 0 .
\label{eq:s_energy}
\end{equation}
Second, closure of the latent trajectory at the end of the melt season,
$\theta(t_e)=\theta_f$, applied to Eq.~(\ref{eq:s_solution}), yields the weighted
constraint
\begin{equation}
\int_{t_s}^{t_e} \mathcal{R}(t) e^{\lambda t}\,dt
=
 \int_{t_s}^{t_e} Q(t) e^{\lambda t}\,dt
\equiv K .
\label{eq:s_boundary}
\end{equation}
These two linear constraints are independent and together define the admissible
class of latent trajectories compatible with both energy conservation and
boundary closure.

\subsection{Minimum Structural Perturbation Principle}

Among all admissible residuals, we select the physically relevant solution by
minimizing the structural deviation from ideal linear relaxation. This is
formalized by minimizing the quadratic functional
\begin{equation}
J[\mathcal{R}]
= \int_{t_s}^{t_e} \mathcal{R}(t)^2\,dt ,
\label{eq:s_action}
\end{equation}
subject to the constraints (\ref{eq:s_energy}) and (\ref{eq:s_boundary}).

Introducing Lagrange multipliers $\mu_1$ and $\mu_2$, we define the augmented
functional
\begin{equation}
\mathcal{L}
= \int_{t_s}^{t_e} \mathcal{R}^2\,dt
+ \mu_1 \int_{t_s}^{t_e} \mathcal{R}\,dt
+ \mu_2 \left(
\int_{t_s}^{t_e} \mathcal{R} e^{\lambda t}\,dt - K
\right).
\end{equation}
Stationarity with respect to variations in $\mathcal{R}(t)$ yields
\begin{equation}
\frac{\delta \mathcal{L}}{\delta \mathcal{R}}
= 2\mathcal{R}(t) + \mu_1 + \mu_2 e^{\lambda t} = 0 ,
\end{equation}
from which the optimal residual takes the form
\begin{equation}
\mathcal{R}^*(t) = \alpha + \beta e^{\lambda t},
\label{eq:s_Ropt}
\end{equation}
where $\alpha=-\mu_1/2$ and $\beta=-\mu_2/2$.
This result reflects the general property that the minimum-norm solution lies in
the subspace spanned by the constraint kernels.

\subsection{Determination of Coefficients}

Substituting Eq.~(\ref{eq:s_Ropt}) into the two admissibility constraints yields
the linear system
\begin{equation}
\begin{pmatrix}
\displaystyle \int_{t_s}^{t_e} dt &
\displaystyle \int_{t_s}^{t_e} e^{\lambda t}\,dt \\[2mm]
\displaystyle \int_{t_s}^{t_e} e^{\lambda t}\,dt &
\displaystyle \int_{t_s}^{t_e} e^{2\lambda t}\,dt
\end{pmatrix}
\begin{pmatrix}
\alpha \\ \beta
\end{pmatrix}
=
\begin{pmatrix}
0 \\ K
\end{pmatrix},
\label{eq:s_matrix}
\end{equation}
where $K$ is defined by Eq.~(\ref{eq:s_boundary}).
All integrals are analytic, ensuring a smooth and numerically stable solution for
$\mathcal{R}^*(t)$ and $\theta(t)$ for any prescribed $\tau$.

\subsection{Selection of the Dissipation Timescale}

While the admissibility constraints are satisfied for any $\tau$, the physically
relevant dissipation timescale is uniquely selected by the principle of minimum
structural perturbation. For each candidate $\tau$, the reconstructed residual
$\mathcal{R}^*(t;\tau)$ yields the objective function
\begin{equation}
f(\tau)
= \int_{t_s}^{t_e} [\mathcal{R}^*(t;\tau)]^2\,dt .
\end{equation}
The optimal timescale is obtained as
\begin{equation}
\tau^* = \arg\min_{\tau \in [\tau_{\min},\tau_{\max}]} f(\tau),
\end{equation}
ensuring that $\tau$ reflects the intrinsic efficiency of surface energy
dissipation rather than an imposed tuning parameter.

\subsection{Implementation}

The reconstruction procedure is implemented as follows:
\begin{enumerate}
\item Identify the melt interval $[t_s,t_e]$ from the observed forcing.
\item For a trial $\tau$, compute $\lambda=(\rho C_s \tau)^{-1}$.
\item Solve Eq.~(\ref{eq:s_matrix}) for $\alpha$ and $\beta$.
\item Evaluate $f(\tau)$ and iterate to determine $\tau^*$.
\item Reconstruct $\theta(t)$ from Eq.~(\ref{eq:s_solution}) using
$\mathcal{R}^*(t)$.
\end{enumerate}

\end{document}